# MINING THE RELATIONSHIP BETWEEN DEMOGRAPHIC VARIABLES AND BRAND ASSOCIATIONS

Ajayeb Abu Daabes[1] and Faten Kharbat[2]

[1]Emirates College of Technology, Abu Dhabi, UAE
Ajayeb.daabes@ect.ac.ae
[2]Al Ain University of Science and Technology, Abu Dhabi, UAE
Faten.kharbat@aau.ac.ae

## ABSTRACT

*This research aims to mine the relationship between demographic variables and brand associations, and study the relative importance of these variables. The study is conducted on fast-food restaurant brands chains in Jordan. The result ranks and evaluates the demographic variables in relation with the brand associations for the selected sample. Discovering brand associations according to demographic variables reveals many facts and linkages in the context of Jordanian culture. Suggestions are given accordingly for marketers to benefits from to build their strategies and direct their decisions. Also, data mining technique used in this study reflects a new trend for studying and analyzing marketing samples.*

## 1. INTRODUCTION

Exploring and analyzing characteristics of consumers can help marketers to target the right customers in the right way by the right tools. Understanding the relationship between demographic characteristics and brand associations has many indicators that reflect the customer behaviour from one side, and how to tailor marketing mix to fit and satisfy customers' needs and preferences from the other side.

According to Aaker [1], brand associations are "the category of a brand's assets and liabilities that include anything ``linked" in memory to a brand". Keller [2] defines brand associations as "informational nodes linked to the brand node in memory and contain the meaning of the brand for consumers. Associations come in all forms and may reflect characteristics of the product or aspects independent of the product".

Scholars variously described brand image as "the perceptions and beliefs held by consumers, as reflected in the associations held in consumer memory" [3]. Several efforts have studied specific elements of brand associations; for example, [1], [4], [5], [6], [7]. Brand associations form brand image which is composed of anything deeply placed in customer's mind toward the brand. Associations may be implicit or explicit perceptions with a brand name through information from commercial/ non-commercial sources (e.g., word of mouth, media, country of origin, channel of distribution, and many other sources). Therefore, the marketers should manage well these sources to influence the consumers' image to be associated strongly with something positive and unique.

The huge amount of data with complicated relations necessitates using other techniques, such as data mining, that are concerned with formulating the process of generalization as a search through possible hypothesis instead of focusing on testing exist hypothesis which is applied by conventional techniques. Data mining is defined to be:" the process of discovering meaningful new correlations, patterns and trends by sifting through large amounts of data stored in repositories, using pat-tern recognition technologies as well as statistical and mathematical techniques" [9]. Accordingly, this study aims to achieve the following goals:

- Study the relationship between brand associations and demographic variables through data mining techniques.
- Study demographic variables according to their weight on brand associations.
- Recommend some suggestions for marketers to direct marketing strategies for their target markets.

This study is organized into five sections. Section 2 overviews some of the recent related work for the brand associations and customer behaviours. Section 3 describes the methodology used in this study. Sections 4 and 5 illustrate the results and conclusions for the study along with some future directions.

## 2. RELATED WORK

The relationship between brand associations and customer characteristics could be measured by different conventional tools. Keller [2] defines quantitative research techniques to be "a means to better approximate the breadth and depth of brand awareness, the strength, favourability, and uniqueness of brand associations, the favourability of brand responses, and the nature of brand relationships". Quantities trend is very useful in concluding summaries which represent various types of scale questions. They have been used as "a primary ingredient in tracking studies that monitor brand knowledge structures of consumers over time" [2]. Qualitative researches techniques [10], in the other hand, "are a means of identify possible brand association" [2]. They have unstructured nature and used to provide an in-depth of what brands and products mean to consumers [11].

Understanding the relationship between customer characteristics and brand associations has been an interesting area of research [15, 16]. Several studies conducted recently aimed to study many elements of consumer brand associations throughout qualitative and quantitative researches using conventional tools to analyze data.

Srivastava [13] conducted a descriptive study to ascertain the perception change in the brand identity of a well known soap brand, brand personality, awareness levels, usage pattern, promotional campaigns and the factors influencing the buying behaviour of the consumers. The study observed that most of the respondents identified the soap clearly as a family/male soap. The study focused on the main issues affecting the soap's brand identity, the markets it has entered, the positioning strategies it has adopted, the celebrities used in endorsing the product, the pricing techniques used.

Another study was conducted by Ménde, Oubiña, & Rubio [14] aims to analyze the relative importance of brand-packaging, price and taste in the formation of brand preference for manufacturer and store brands in food product categories. The study used a blind taste test of the product using three brands (two manufacturer brands and one store brand). It used con-joint analysis to analyze the influence of the intrinsic cue (taste) and the extrinsic cues (price and brand-packaging) on consumers' preference for manufacturer and store brands. The results show that not knowing the brand to which the taste tested belongs, leads consumers in general to order their preferences fundamentally by taste. However, the results differ by product category and consumer segment analyzed. Consumers who evaluate the taste of store brands as better change their preferences more when they know which brand belongs to which taste. Further, the change in preference when consumers know the brand-taste correspondence is clearly greater in the most differentiated category. Store brands are only evaluated more favourably than manufacturer brands in price.

Alimen & Cerit [15] analyzed via exploratory study the impacts of gender, field of education, and having consumed the brand, on consumers' brand knowledge. The survey was con-ducted by using convenience sampling over Turkish university students. The study discovered significant differences with respect to usage, gender, and departments. The study revealed that students in departments more related to fashion and female students have more knowledge about the tested brands. Also, "the findings demonstrate that consumption of a brand increases both brand awareness and brand image" [15].

Alexandris et al. [17] focused on the exploratory factor analysis and revealed eight brand association factors: popularity, management, logo, escape, vicarious achievement, nostalgia, pride and affect. The study was conducted over 165 participants of a managed-owned fitness club. Two questionnaires were designed of 25 questions each, one to measure brand associations, and the other to measure service quality. It was found that five of the eight brand associations (escape, nostalgia, pride, logo, and affect) significantly contributed to the prediction of loyalty. Also, it was revealed that the service quality dimensions predicted significant amount of variances in all the eight brand associations.

## 3. RESEARCH METHODOLOGY

This study applies some of the well known data mining techniques to study the relative importance of demographic variables and the favourable bonds with brands associations for fast food restaurants' brand. In this research we aim to study some of restaurants' brand associations in relation with customers' demographic variables. The following restaurants' brand associations were studied:

1. *Brand Elements* which are defined as the name or logo or any other symbols represent the restaurant.

2. *Emotions* that reflect feelings that are carried toward the brand.

3. *Trust* that means the level of credibility the brand stands for.

4. *Distribution Channels* that reflect availability and convenience for the customer.

5. *Brand Quality* that reflects perceived quality from customer.

6. *Country of Origin* that represents the perceived image the customer has toward the country of origin.

7. *Price* that indicates the level of affordability for the customer.

8. *Loyalty* that represents positive feelings toward the brand that lead to a commitment to re-purchase the brand and continue using it.

Experiments were performed using two well-known data mining techniques via Weka software [18]. Weka is a collection of machine learning algorithms for solving real-world data mining problems [18]. The Weka workbench [19] contains a collection of visualization tools and algorithms for data analysis and predictive modelling, together with graphical user inter-faces for easy access to this functionality. The first data mining technique is the X-Mean Clustering Algorithm which is an extended K-Means Clustering algorithm [20] by an improve-structure part in which the centres are challenged to be split in its region. The decision between the records of each centre and itself is done comparing the BIC-values of the two structures [21]. The second data mining algorithm used in this research is the Ranker Algorithm which "ranks attributes by their individual evaluations." [19]. The algorithm is used in conjunction with the Relief algorithm (ReliefF [23]) to estimate and evaluate the variables. It is also used to detect conditional dependencies between data features. Many researchers have demonstrated that ReliefF algorithm is very useful in different applications and domains, for example [24] and [25]. It is usually used to deal with data that is noisy, incomplete, or/and has more than two classes.

The questionnaire was developed to collect the primary data needed for this study; it consists of two main parts: the first part includes questions about demographic data of the respondent, which are: gender, marital status, educational degree, age, and income level. Table 1 illustrates the variables and their categories along with their frequencies.

The second part of the questionnaire includes eight variables. 49 statements were developed to cover all the variables. Likert scale in five levels was applied. The variables: (1) Brand elements were measured in paragraphs 1-5. (2) Brand quality was measured in paragraphs 6-18, (3) Brand trust was measured in paragraphs 19-22, (4) Emotions were measured in paragraphs 23-27, (5) Distribution channels were measured in paragraphs 28-33, (6) Price was measured in paragraphs 34-38, (7) Country of origin was measured in paragraphs 38-43, and (8) Loyalty was measured in paragraphs 44-49.

Table 1 : Summary of Demographic Variables

| Demographic Variables | Category | Frequency | % |
|---|---|---|---|
| Gender | Male | 493 | 61 |
| | Female | 319 | 39 |
| Age | Less than 20 years | 203 | 25 |
| | 30-20 | 435 | 54 |
| | 40-31 | 74 | 9 |
| | 50-41 | 43 | 5 |
| | More than 50 years | 57 | 7 |
| Marital status | Single | 637 | 78 |
| | Married | 135 | 17 |
| | Divorcee | 40 | 5 |
| Educational degree | Less than High School | 49 | 6 |
| | High School | 142 | 18 |
| | Diploma | 30 | 4 |
| | Bachelor | 474 | 58 |
| | Graduate | 117 | 14 |
| Income level | Less than 500 JD | 118 | 15 |
| | 1000-500 | 262 | 32 |
| | 1500-1001 | 169 | 21 |
| | 2000-1501 | 90 | 11 |
| | More than 2000JD | 173 | 21 |

A total of 1067 survey questionnaires were distributed via convenience sample covering fast food restaurant brand chains' clients in the capital of Jordan, Amman. From this, 889 questionnaires were returned and 77 questionnaires were dropped from the analysis due to incomplete answers, with a final sample of 812 respondents representing a response rate of 76%. Briefly, the respondents consisted of 61% (n=493) males, about 54% the age group of 20-30 years old, about 78% were single, and more than half had university education. All in-come ranges were covered in the questionnaire, but about 32% of the respondents had an estimated monthly income between 500-1000 JD. A summary of demographic variables is presented above in Table 1.

## 4. RESULTS

Our work has two main results. First, we present the findings for the X-Means algorithm over each demographic variable separately. Next, we evaluate the variables to discover their ranking which gives us indication about their importance to the brand associations.

After applying X-Means algorithm over the sample, five figures are produced to reflect each demographic variable separately in relation with brand associations using excel line chart type as shown in Figure 1.

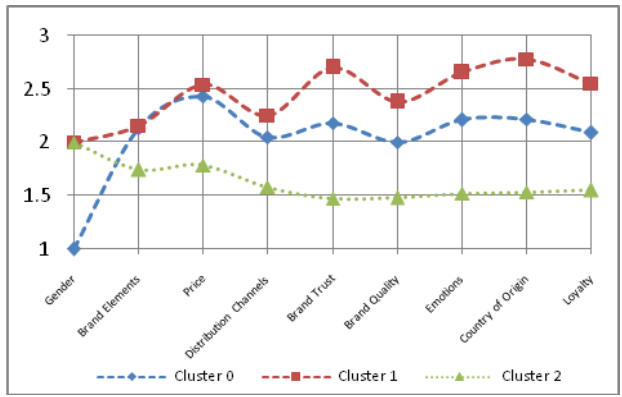

a) Clusters based on Gender

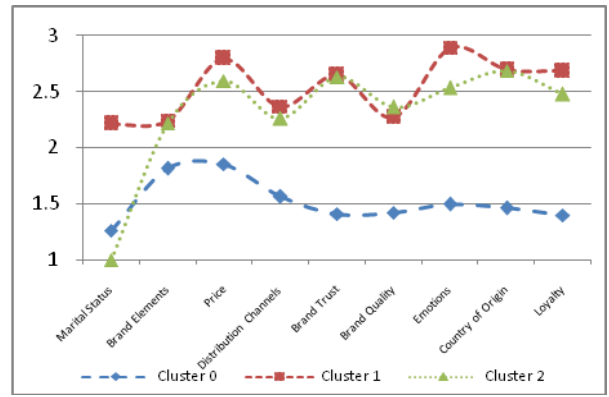

b) Clusters based on Marital Status

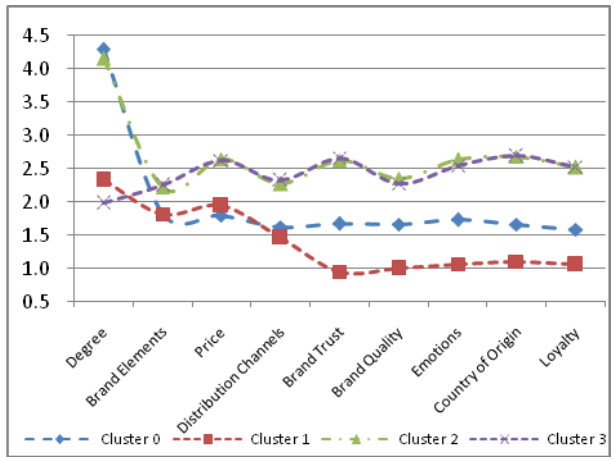

c) Clusters based on Degree Level

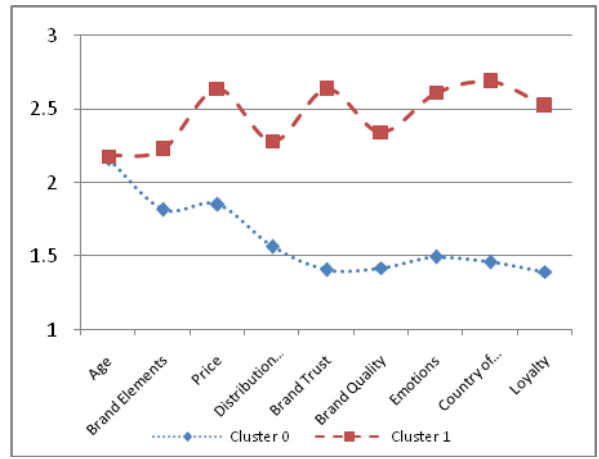

d) Clusters based on Age

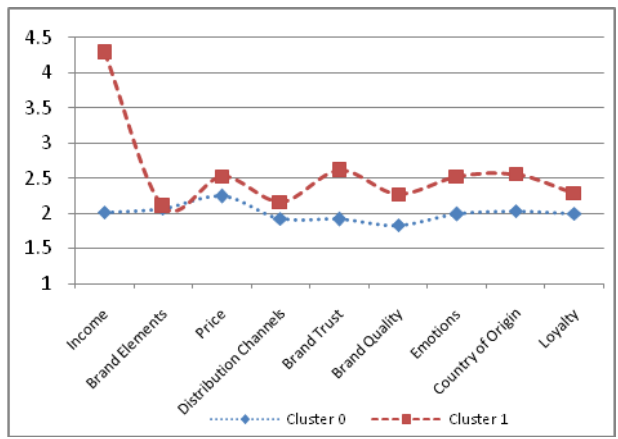

e) Clusters based on Income

Figure 1: The Clusters Based on the Demographic Variables

Figure 1(a) reveals that there are positive associations between the customers' gender and the fast food restaurants' brand. The figure shows that the sample is divided into three groups; the first and the second groups are very similar in their trend. They show neutral and favourable relations with brand associations between customers and fast food restaurants brands; that is, most of the associations have a centre around or above 2. The first group is for the males customers and the second group is for part of the females' customers. The third group is for the other part of the females' customer which shows different trend towards brand associations; it reflects weak relationship with brand associations.

Figure 1(b) reveals that there are positive associations between the customers' marital status and the fast food restaurants' brand associations. The figure shows that the sample is divided into three groups; married customer has good and favourable relationship with fast food restaurants brand along with part of the singles customers. On the other hand there is a segment of single customers that have below average relationship with brand associations.

Figure 1(c) reveals that there are positive associations between the customers' educational degree and the fast food restaurants' brand associations. Although the figure shows four different groups, the sample trend seems to have three different tendencies. Graduate and bachelors' educational degree have positive and favourable relationship with fast food restaurants brand associations along with segment of high secondary schools customers. The remaining two groups are for diploma and segment of high secondary schools customers which have below average relationship with brand associations.

Figure 1(d) reveals that there are positive associations between the customers' age and the fast food restaurants' brand associations. The figure shows that one group of the sample has significant relationship, which is the customers between 20-30 years old, and brand association. This group is divided into two different clusters. One of them has positive and favourable relationship with fast food restaurants brand and the other one has weak relationship.

Figure 1(e) reveals that there are positive associations between the customers' income and the fast food restaurants' brand associations. The figure shows two groups that have significant trend; customers' incomes' between 500 and 1000JD, and between 1501 and 2000JD. Customers from the two categories reflect positive and favourable relationship with fast food restaurants' brand.

The second part for our result is the output of the Ranking and Relief algorithm which is a weight between −1 and 1 for each variable; the more positive weights indicate more predictive attributes. Table 2 illustrates the demographic variables and their weight based on the used algorithm. The result reveals that the Income variable is the most important indicator for the current sample on the brand associations. On the other hand, the age variable has shown to have the least effect.

Table 2: The Demographic Variables' Weights Using the Ranking and Relief algorithm

| Weight | DV |
|---|---|
| -0.01017 | Age |
| -0.0075 | Sex |
| -0.0036 | Degree |
| -0.00268 | Material |
| 0.00895 | Income |

(More predictive ↓)

## 5. DISCUSSION

The conclusions of this study split into two parts; the first one is that using the X-Means algorithm over all the demographic variables help discovering brand associations and suggest some relations between the variables and brand associations. Findings reflect neutral and favourable relationship between demographic variables and brand associations. These results support the "variety seeking buying behaviour [22]" type: which occurs in situations "of low consumer involvement but significant perceived brand differences. In such cases consumers often do a lot of brand switching" [22]. This indicates that there is no real brand loyalty; hence, further effort should be conducted in future research to distinguish between associations. However, some demographic variables show to have a unique trend toward brand associations; for example, married customers seem to have positive trend towards brand associations, some of the females' customers tend to have negative brand associations, and the customers from the middle age seem to have two different trends with the brand associations.

All males group and a segment of females have favourable relationship with brand associations while the remaining segment of females has weak relationship. This result refers to the culture prevailed in Jordanian society which allows males to move more freely, and that definitely helps them to build some associations with their brands.

Being single has strong effect on the brand associations. This result seems to match the Jordanian social customs which encourage family relations and home cooking meals. Singles studying or working in Amman may be from another city and hence their bonds with the restaurants are strong. In general, singles have less responsibility towards their families and strong bonds with their friends which encourage them to use the restaurant services more often.

Graduate and bachelor's customers have significant relationship with brand associations; this result aligns with the level of awareness within these categories that help to build and create positive relationships with brands.

Both married and singles customers have significant relationship with brand association except a segment of singles customers. This result seems to be logic since married customers have a lot of standards to make purchasing decisions. While singles have different trends, some of them has positive relations with brand associations, and the other one has weak relations which indicates that this segment don't care of about fast food brand associations.

For the income element, there are no differences between the two clusters, but the figure shows that both clusters should be in high income range to build such an associations. This result is logic since the income level in Jordan (Amman in particular) is low, and the visits to restaurants may be considered a complementary expense for low incomes. For that the output of the Relief algorithm reflects that the income is the most effective factor in comparison with other demographic variables in this study.

Customers (20-30) years old seemed to have significant relationships rather than other categories. This result aligns with nature of Jordanian society in which the majority of this category is unmarried therefore they visit fast food restaurants frequently and building favourable relationships with these brands.

It recommends a practical technique that can help marketer to forming their strategies and decisions using well known data mining techniques instead of pure traditional statistical techniques. Data mining techniques are used to discover implicit interesting patterns and relations within the collected data using machine learning, pattern recognition, statistics, databases, and algorithms. This study opens the door for using such techniques in the field of branding. Further investigations may illustrate a comparison study between the results using pure statistical methods and data mining techniques.

The second part is highlighting the brand associations between customers and fast food restaurants brands chains in Jordan. The result reveals that there are neutral and favourable associations, and this result is reasonable since the switching cost is low and the customers want to check variety of services. The study shows that some demographic variables have significant impact on brand associations, such as, income, marital status, and educational degree. Therefore, further efforts have taken place to examine the strengths of the suggested associations via ranking technique to rank the variables according to their impact on brand associations as shown above.

## 6. CONCLUSIONS AND FUTURE WORK

The conclusions of this study split into two parts; the first one is that using some data mining techniques, such as X-Means algorithm, over the demographic variables help discovering brand associations and suggest relations between these variables. The findings reflect neutral and favourable relationship between demographic variables and brand associations. These results support the "variety seeking buying behaviour" type: which occurs in situations "of low consumer involvement but significant perceived brand differences. In such cases consumers often do a lot of brand switching" [22]. This indicates that there is no real brand loyalty for the restaurants' brands; hence, further effort should be conducted in future research to distinguish between associations. However, some demographic variables show to have a unique trend toward brand associations; for example, income, marital status, and educational degree have significant differences between customers and brand associations.

The second part of the conclusion relies on predicting significant targets for marketers that is derived from the previous results. In general all customers hold neutral associations with restaurants' brand. Hence, this fact creates a lot of pressure on marketers to switch customer form neutral status to loyalty status.

Segments of the female group have shown weak associations with restaurants' brand; this requires marketers to attract this segment through their brand strategies. Also, category of customers between 20 and 30 years old presents significant relation with the associations, which indicates positive attitudes towards fast food brands. Marketers are advised to focus on this category to make them more loyal and to attract other ages' segments.

High income customers seem to have a good trend towards brand associations; however, low income levels have negative attitude toward fast food brands. Marketers should persuade other categories with their pricing strategies.

High educational customers who have significant relations with restaurants' brand reflect that lower educational level customers may have an ignorance relation, which means that customers from this segment does not care about any associations!

Married customers and singles both are targeted for fast food restaurants; marketers shouldn't ignore any segment unless they scan the culture and their trends.

Finally, this research recommends a practical technique that can help marketer to form their strategies and decisions using well known data mining techniques instead of pure conventional techniques. This study opens the door for using such techniques in the field of branding. Further investigations may take place to compare this study with others from different culture and countries.

**Authors**

**Ajayeb Abu Daabes** is an Assistant Professor in Marketing at Emirates College of Technology, Abu Dhabi, UAE. She holds PhD degree in Marketing/Branding from Amman Arab University, Jordan, in 2009. Her Master degree was in MBA/Marketing in 2002 from Jordan University. Her main research interest is customer behavior, service marketing, brand management, and relationship marketing.

**Faten Kharbat** is an Assistant Professor in Artificial Intelligence at the Al Ain University, Abu Dhabi Campus, UAE. She holds PhD degree in computer science from the University of the West of England, UK, in 2006. Her main research interest is Learning Classifier Systems, knowledge based systems, applying data mining techniques to marketing, website design, and recently was involved in e-learning and quality of higher education.